\documentclass[%
groupedaddress,
unsortedaddress,
nofootinbib,
 amsmath,amssymb,
 aps,
twocolumn,
]{revtex4-1}

\usepackage[utf8]{inputenc}
\usepackage[english]{babel}
\usepackage{titlesec}
\usepackage{etoolbox}

\usepackage{amsmath}
\usepackage{amssymb} 
\usepackage{booktabs}

\usepackage{graphicx} 
\usepackage{xcolor}
\usepackage{natbib}
\usepackage{braket}
\usepackage{float}
\usepackage[hidelinks]{hyperref}
\usepackage{cleveref}

\usepackage[normalem]{ulem}
\newcommand{\stkout}[1]{\ifmmode\text{\sout{\ensuremath{#1}}}\else\sout{#1}\fi}

\usepackage[all]{nowidow}

\usepackage{newfloat}
\DeclareFloatingEnvironment[
    name=Algorithm
]{algorithm}

\begin{document}

\preprint{APS/123-QED}

\title{Revisiting Shooting Point Monte Carlo Methods for Transition Path Sampling}
\date{\today}

\author{Sebastian Falkner 
}
\affiliation{Faculty of Physics, University of Vienna, 1090 Vienna, Austria}
\affiliation{Institute of Physics, University of Augsburg, Universit\"atsstraße 1, 86159 Augsburg, Germany.}

\author{Alessandro Coretti 
}
\affiliation{Faculty of Physics, University of Vienna, 1090 Vienna, Austria.}

\author{Baron Peters 
}
\affiliation{Chemical and Biomolecular Engineering, University of Illinois at Urbana-Champaign, Urbana, Illinois 61801, USA}
\affiliation{Department of Chemistry, University of Illinois at Urbana-Champaign, Urbana, Illinois 61801, USA}

\author{Peter G. Bolhuis 
}
\affiliation{Van ’t Hoff Institute for Molecular Sciences, University of Amsterdam, PO Box 94157, 1090GD Amsterdam, The Netherlands}

\author{Christoph Dellago 
}
\email{christoph.dellago@univie.ac.at}
\affiliation{Faculty of Physics, University of Vienna, 1090 Vienna, Austria}
\affiliation{Research Platform on Accelerating Photoreaction Discovery (ViRAPID), University of Vienna, 1090 Vienna, Austria}

\begin{abstract}

Rare event sampling algorithms are essential for understanding processes that occur infrequently on the molecular scale, yet they are important for the long-time dynamics of complex molecular systems. One of these algorithms, transition path sampling (TPS), has become a standard technique to study such rare processes since no prior knowledge on the transition region is required. Most TPS methods generate new trajectories from old trajectories by selecting a point along the old trajectory, modifying its momentum in some way, and then ``shooting'' a new trajectory by integrating forward and backward in time. In some procedures, the shooting point is selected independently for each trial move, but in others, the shooting point evolves from one path to the next so that successive shooting points are related to each other. To account for this memory effect, we introduce a theoretical framework based on an extended ensemble that includes both paths and shooting indices. We derive appropriate acceptance rules for various path sampling algorithms in this extended formalism, ensuring the correct sampling of the transition path ensemble. Our framework reveals the need for amended acceptance criteria in the flexible-length aimless shooting and spring shooting methods.

\end{abstract}

\maketitle

\section{Introduction}
Transition path sampling (TPS) is a simulation method that yields unbiased insights into the mechanisms and kinetics of rare events occurring in complex molecular systems. It is applicable to a wide range of processes, including nucleation~\cite{Menzl2016}, biomolecular reorganization~\cite{Okazaki2019,Juraszek2006} and chemical reactions~\cite{Basner2005,Geissler2001}. Unlike many enhanced sampling techniques, TPS does not require detailed a priori knowledge about the process of interest, other than a definition of the start and end states of the process~\cite{Bolhuis2002}. TPS samples from a distribution of unbiased dynamical trajectories with start and end points located in these predefined (meta)stable states. This ensemble of transition paths is generated sequentially by employing a Markov chain Monte Carlo framework~\cite{Bolhuis2002}. At the core of many path sampling schemes is the so-called shooting move, in which a new path is generated by the integration of the equations of motion (forward and backward in time) starting from a preselected configuration taken from an existing transition path, known as  
the \emph{shooting point}. Valid reactive paths that connect the start and end states are accepted or rejected in a Metropolis step. Standard TPS has successfully been used on a wide variety of systems and processes~\cite{Bolhuis2015,Bolhuis2021}.

Several methods have been proposed to increase the sampling efficiency of TPS by incorporating prior knowledge of the system into the selection of the shooting point using collective variables, such as biased TPS~\cite{Bolhuis2002,Juraszek2006} or shooting range TPS~\cite{Jung2017}. Likewise, artificial intelligence-assisted sampling can increase the sampling efficiency significantly~\cite{Jung2023,Falkner2023} and enable efficient rate calculations~\cite{Lazzeri2023}. While methods involving prior knowledge about the reaction are useful in many cases, such knowledge is often not available. For such cases, aimless shooting~\cite{Peters2006,Mullen2015,Peters2010} and spring shooting~\cite{Brotzakis2016} were introduced as system-agnostic, yet  efficient, alternatives~\cite{Beckham2011,Joswiak2018,Arjun2021,Janos2020}.
While in most other TPS procedures the shooting point is selected freshly from the current path for each trial move, in the aimless and spring shooting methods, the new shooting point is chosen based on knowledge of the previous shooting point. As a result, the shooting point evolves in a certain way from one path to the next. A key concept in these methods is that incorporating memory of the shooting point allows the algorithm to focus automatically on the barrier region, thereby increasing the acceptance probability of pathways.  

In this study, we revisit several history-based shooting point Monte Carlo methods, including flexible-length aimless shooting and spring shooting. We find that the particular procedures to select shooting points in these methods lead to violations of detailed balance, resulting in an improperly sampled path ensemble. This issue arises because the generation of the new shooting point depends on information on the previous shooting point.

Our objective here is to develop a general formalism that accounts for the evolving shooting point in order to restore detailed balance and ensure correct sampling of the desired path distribution. This formalism makes use of an extended state space that includes shooting points in addition to trajectories. We develop the general formalism and apply it to the case of the aimless and spring shooting algorithms. In particular, we show that the performance of the fixed-length versions of these methods depends strongly on the total path length. Furthermore, we find a detailed balance-related issue in the original spring shooting method~\cite{Brotzakis2016}, which we correct in the new formalism. Unfortunately, one of the consequences of treating the algorithms correctly in the extended space formalism is that they are not behaving as efficiently as concluded in the original papers.

The remainder of the paper is organized as follows: first, we introduce the extended space of paths and shooting indices, necessary to capture the evolution of the shooting point. Next, we derive detailed balance relations for Monte Carlo moves in this extended space and apply them to regular TPS, aimless shooting and spring shooting. This formalism leads to corrections of the original flexible-length aimless and spring shooting algorithms, which we then validate by performing path sampling in a one-dimensional test system.

\section{Theory}

\subsection{Fixed- and Flexible-length Path Ensembles}


Assuming two (meta)stable states A and B, we consider an ensemble of discretized trajectories $X = \{x_1, x_2, ... x_{L}\}$ that connect A and B, where each time slice or frame $x$ contains the coordinates (and possibly momenta) of all particles in the system. We define a probability density in the ensemble of transition paths as $P_{\text{AB}}(X)$. For instance, a path distribution where all paths are of fixed length $L$ can be defined as
\begin{align}
    P_{\text{AB}}(X) = \frac{1}{Z_{\text{AB}}} h_{\text{A}}(x_1) h_{\text{B}}(x_L) \rho(x_1) \prod_{i=1}^{L-1} p(x_i \to x_{i+1})\, ,
\end{align}
where $h_{\text{A}}(x)$ and $h_{\text{B}}(x)$ are the characteristic functions of the states A and B (unity when $x\in \text{A},\text{B}$, zero otherwise), $\rho(x_1)$ is the stationary distribution generated by the underlying dynamics (e.g.,~the Boltzmann distribution) and $p(x_i \to x_{i+1})$ is the short time transition probability from $x_i$ to $x_{i+1}$. The partition function $Z_{\text{AB}}$ normalizes the distribution.

Another commonly used definition is the flexible-length ensemble, in which pathways have different lengths $L(X)$. These paths are required to have the start and endpoint in A and B, respectively, while all other points in between should lie outside of A and B:
\begin{align}
    P_{\text{AB}}(X) \propto H_{\text{AB}}(X) \rho(x_1) \prod_{i=1}^{L(X)-1} p(x_i \to x_{i+1})\, .
\end{align}
Here, $H_{\text{AB}}(X)$ is unity if the path $X$ fulfills all requirements to be considered reactive and zero otherwise, as specified below, i.e.~$x_1\in \mathrm{A}, x_L\in \mathrm{B}$.

\subsection{Extended Space of Paths and Shooting Indices}

In path sampling methods like aimless shooting and spring shooting, information on the previous shooting point is used to generate the new shooting point. To take this information into account, we denote with $k$ the index of a (configuration) point $x_k$ on the path $X$. Depending on the sampling scheme, this point can either be a shooting point directly or a point that is used to generate the next shooting point. The exact interpretation of $k$ will become clear later when we discuss several path sampling algorithms (see, for example,~\cref{sec:TPS,sec:aimless_fixed,sec:aimless,sec:spring}). We refer to $k$ as the {\em shooting index} in both cases since, even in the second case, the move is initiated from $k$. Common for both cases is that the index $k$, in some form or the other, evolves from trial to trial. 

To track the evolution of $X$ and $k$ at the same time in a Monte Carlo procedure while preserving the Markovianity of the process, we define an extended space that includes both the paths and the shooting indices. Each point $Y$ in this extended space is composed of a path $X$ and a shooting index $k$,
\begin{align}
Y = (X, k)\, .
\end{align}
In each step of the Monte Carlo procedure, a new path $X^{\text{n}}$ is generated together with a new shooting index $k^{\text{n}}$:
\begin{align}
(X^{\text{o}}, k^{\text{o}}) \to (X^{\text{n}}, k^{\text{n}})\, ,
\end{align}
with a particular generation probability 
\begin{align}
p_{\text{gen}}\bigl[ (X^{\text{o}}, k^{\text{o}}) \to (X^{\text{n}}, k^{\text{n}}) \bigr]\, ,
\end{align}
that depends on the details of the generation algorithm. The new path and shooting index are then accepted with acceptance probability $p_{\text{acc}}\bigl[ (X^{\text{o}}, k^{\text{o}}) \to (X^{\text{n}}, k^{\text{n}}) \bigr]$. If $(X^{\text{n}}, k^{\text{n}})$ is rejected, the old path and shooting index are counted again in the path or index ensembles and the corresponding averages.

To derive an acceptance probability for the Monte Carlo move, the joint probability density $P_\mathrm{AB}(X, k)$ must be specified. Since our goal is to sample the transition path ensemble $P_{\text{AB}}(X)$, we require that marginalizing $P_\mathrm{AB}(X, k)$ with respect to $k$ yields $P_{\text{AB}}(X)$,
\begin{align}
\sum_k P_\mathrm{AB}(X, k) = P_{\text{AB}}(X)\, .
\end{align}
This condition is naturally fulfilled by defining $P_\mathrm{AB}(X, k)$ via the conditional probability $p(k | X)$, which has been described previously in the context of auxiliary variable methods~\cite{Higdon1998}:
\begin{align}
P_\mathrm{AB}(X, k) = P_{\text{AB}}(X) p(k | X)\, .
\end{align}
Due to the normalization of the conditional probability, $\sum_k p(k|X) = 1$, the sum over all $k$ is guaranteed to lead to $P_{\text{AB}}(X)$ as marginal distribution. Apart from this condition, there is total freedom in setting the concrete form of $p(k | X)$, the target distribution for the shooting index $k$.

\subsection{Shooting Point Monte Carlo}

The goal is now to construct a Markov chain Monte Carlo procedure for sampling the joint probability $P_\mathrm{AB}(X, k) = P_{\text{AB}}(X) p(k | X)$. In the subspace of the trajectories $X$ this procedure then samples the desired transition path ensemble $ P_{\text{AB}}(X)$. Imposing detailed balance
\begin{align}
    &P_{\text{AB}}(X^{\text{o}})  p(k^{\text{o}} | X^{\text{o}}) p_{\text{gen}}\bigl[ (X^{\text{o}}, k^{\text{o}}) \to (X^{\text{n}}, k^{\text{n}}) \bigr] \notag \\
    & \times p_{\text{acc}}\bigl[ (X^{\text{o}}, k^{\text{o}}) \to (X^{\text{n}}, k^{\text{n}}) \bigr] =\notag \\
   &P_{\text{AB}}(X^{\text{n}})  p(k^{\text{n}} | X^{\text{n}}) p_{\text{gen}}\bigl[ (X^{\text{n}}, k^{\text{n}}) \to (X^{\text{o}}, k^{\text{o}}) \bigr]  \notag \\
    &\times p_{\text{acc}}\bigl[ (X^{\text{n}}, k^{\text{n}}) \to (X^{\text{o}}, k^{\text{o}}) \bigr]
\end{align}
yields
\begin{align}
    \label{eq:acceptance_full}
    &\frac{p_{\text{acc}}\bigl[ (X^{\text{o}}, k^{\text{o}}) \to (X^{\text{n}}, k^{\text{n}}) \bigr] }{p_{\text{acc}}\bigl[ (X^{\text{n}}, k^{\text{n}}) \to (X^{\text{o}}, k^{\text{o}}) \bigr]} =\notag \\
    &\frac{P_{\text{AB}}(X^{\text{n}})  p(k^{\text{n}} | X^{\text{n}}) p_{\text{gen}}\bigl[ (X^{\text{n}}, k^{\text{n}}) \to (X^{\text{o}}, k^{\text{o}}) \bigr] }{P_{\text{AB}}(X^{\text{o}})  p(k^{\text{o}} | X^{\text{o}}) p_{\text{gen}}\bigl[ (X^{\text{o}}, k^{\text{o}}) \to (X^{\text{n}}, k^{\text{n}}) \bigr]}\, ,
\end{align}
which can be satisfied with the Metropolis-Hastings criterion. The generation probability can be further factorized as
\begin{align}
    \label{eq:pgen_factorization}
    p_{\text{gen}}\bigl[ (X^{\text{o}}, k^{\text{o}})& \to (X^{\text{n}}, k^{\text{n}}) \bigr] = \notag \\
    &p_{\text{gen}}( X^{\text{n}} |  X^{\text{o}}, k^{\text{o}})  p_{\text{gen}}( k^{\text{n}} | X^{\text{n}},  X^{\text{o}}, k^{\text{o}})\, ,
\end{align}
to separate the generation probabilities of the new path from the generation probabilities of the new shooting index. This factorization implies that first a new path is generated according to $p_{\text{gen}}( X^{\text{n}} |  X^{\text{o}}, k^{\text{o}})$ based on the old path and the old shooting index. This probability includes all factors resulting from the propagation rules of the underlying dynamics~\cite{Dellago2003c,Bolhuis2002}. Then, the new shooting index is generated according to $p_{\text{gen}}( k^{\text{n}} | X^{\text{n}},  X^{\text{o}}, k^{\text{o}})$ based on the old shooting  index as well as the old and new path. This factor describes the probability to select the new shooting index $k^{\text{n}}$. Another possibility, corresponding to a different factorization, is to first generate the new shooting index and then the new path or further factorize the shooting index generation process.

To further simplify the acceptance probability, we assume that all pathways are generated using shooting moves based on the underlying dynamics, which conserves the equilibrium distribution. Furthermore, we assume that momenta of the shooting point are either unchanged or independently redrawn from the Maxwell-Boltzmann distribution~\cite{Dellago2003c}. This leads to a cancellation of the path probabilities $P_{\text{AB}}(X)$ and the respective generation probabilities, where we factorize $p_{\text{gen}}$ as in Eq.~\ref{eq:pgen_factorization} \cite{Dellago2003c}. For the flexible-length path ensemble this leads to:
\begin{align}
    \label{eq:acceptance_short}
    &\frac{p_{\text{acc}}\bigl[ (X^{\text{o}}, k^{\text{o}}) \to (X^{\text{n}}, k^{\text{n}}) \bigr] }{p_{\text{acc}}\bigl[ (X^{\text{n}}, k^{\text{n}}) \to (X^{\text{o}}, k^{\text{o}}) \bigr]} =  \notag \\
    & H_{\text{AB}}(X^{\text{n}})
    \frac{ p(k^{\text{n}} | X^{\text{n}}) p_{\text{gen}}( k^{\text{o}} | X^{\text{o}},  X^{\text{n}}, k^{\text{n}}) }
    { p(k^{\text{o}} | X^{\text{o}}) p_{\text{gen}}( k^{\text{n}} | X^{\text{n}},  X^{\text{o}}, k^{\text{o}})}\, .
\end{align}
Equation~\eqref{eq:acceptance_short} can be satisfied by applying the Metropolis-Hastings criterion,
\begin{align}
    \label{eq:acceptance_short_hasting}
    &p_{\text{acc}}\bigl[ (X^{\text{o}}, k^{\text{o}}) \to (X^{\text{n}}, k^{\text{n}}) \bigr] =  \notag \\
    & H_{\text{AB}}(X^{\text{n}})
     \min \biggl[ 1, \frac{ p(k^{\text{n}} | X^{\text{n}}) p_{\text{gen}}( k^{\text{o}} | X^{\text{o}},  X^{\text{n}}, k^{\text{n}}) }
    { p(k^{\text{o}} | X^{\text{o}}) p_{\text{gen}}( k^{\text{n}} | X^{\text{n}},  X^{\text{o}}, k^{\text{o}})} \biggr]\, .
\end{align}

Below, we will use this equation to derive appropriate acceptance rules for different TPS-algorithms. 

\subsection{Shooting Point Densities}

To assess the convergence of a path sampling scheme and further understand its inner workings, it is useful to examine the distribution of points on transition paths, which we note by $\rho(x | \text{TP})$ and the distribution of points from which shooting moves are initiated, indicated by $\rho(x|\text{SP})$ in what follows. For pathways distributed according to $P_{\text{AB}}(X)$, $\rho(x | \text{TP})$ is given by
\begin{align}
    \label{eq:path_point_density}
    \rho(x | \text{TP}) &= 
    \frac{\int \text{d}X\, P_{\text{AB}}(X) \sum_{i=1}^{L(X)} \delta(x - x_i) }{\int dx \int \text{d}X\, P_{\text{AB}}(X) \sum_{i=1}^{L(X)} \delta(x - x_i)} \notag\\  
    &=    \frac{\int \text{d}X\, P_{\text{AB}}(X) \sum_{i=1}^{L(X)} \delta(x - x_i) }{\int \text{d}X\, P_{\text{AB}}(X) L(X) } \notag\\   
    &=\frac{1}{\langle L(X) \rangle_\mathrm{TP}} \int \text{d}X\, P_{\text{AB}}(X) \sum_{i=1}^{L(X)} \delta(x - x_i) \, ,
\end{align}
where the delta function probes the occurrence of $x$ at position $x_i$ on a transition path $X$. In the above equation, the notation $\langle \cdots \rangle_\mathrm{TP}$ implies an average over the transition path ensemble $P_{\text{AB}}(X)$. Hence, $\langle L(X) \rangle_\mathrm{TP}$ is the average length of transition pathways. Note that shorter paths contribute fewer points to the distribution $\rho(x | \text{TP})$ defined in Eq.~\eqref{eq:path_point_density}. For low-dimensional systems, this density can be computed numerically in a TPS simulation by histogramming points on the sampled transition pathways. In higher dimensions, it is more practical to consider the distribution $\rho(r | \text{TP})$ of a collective variable $r(x)$ for points on transition pathways. In either case, the distribution is a key indicator for the correct convergence of a path sampling scheme.

\begin{figure*}
    \centering
    \includegraphics[width=0.6\linewidth]{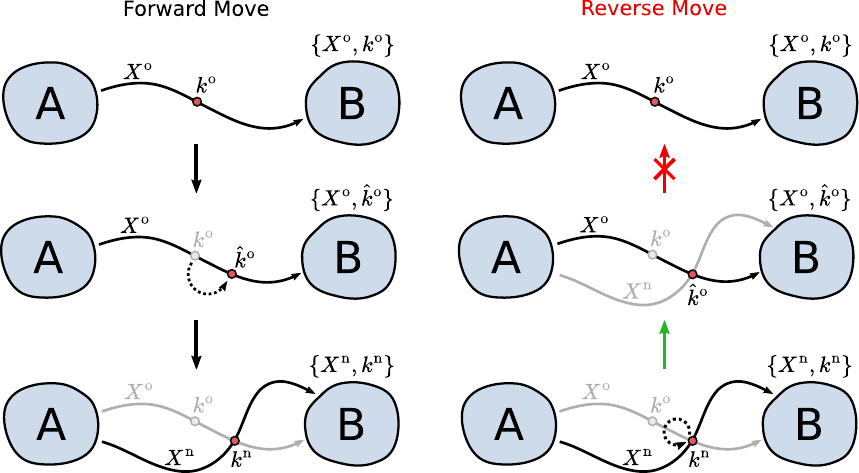}
    \caption{The standard shooting move is irreversible in the extended space framework. Starting from a path $X^{\text{o}}$ and previous shooting index $k^{\text{o}}$ (left), the shooting move is usually divided into two steps:  first, a new index $\hat k^{\text{o}}$ is picked and, second, the new path $X^{\text{n}}$ is generated from the selected point. On the new path the used shooting point has the index $k^{\text{n}}$. Although the reverse move (right) can recover the old path by selecting the same shooting index $k^{\text{n}}$ again, it is not possible to recover the old shooting index (indicated by the red arrow).}
    \label{fig:reversibility_issue}
\end{figure*}

Another distribution that is important to discuss is the distribution of attempted shooting points given that one follows the Monte Carlo scheme described in the previous section. In contrast to $\rho(x | \text{TP})$, which is fully defined by specifying the transition path ensemble $P_{\text{AB}}(X)$, this distribution also depends on the specific protocol employed in the TPS simulation. To obtain a general expression for the distribution of attempted shooting points, we need to take into account that $k$ does not necessarily need to be the shooting index directly, but may also serve to generate an index $\hat k$, from which the shooting move is initiated. Denoting with $p_{\text{sel}}(\hat k | X, k)$ the probability to select $\hat k$ given the current index $k$, the (protocol-dependent) shooting point distribution is given by
\begin{align}
    \label{eq:sp_density_general}
    \rho(x|\text{SP}) = \int \text{d}X\, &P_{\text{AB}}(X) \sum_{k=1}^{L(X)} \biggl[ p(k|X) \notag \\ 
    &\times \sum_{\hat k=1}^{L(X)} p_{\text{sel}}(\hat k | X, k) \delta(x - x_{\hat k}) \biggr]\, .
\end{align}
Note that the shooting point distribution depends both on the imposed distribution $p(k| X)$ of shooting indices as well as on the specific shooting algorithm as encoded in the selection probability $p_{\text{sel}}(\hat k | X, k)$ (and, of course, also on $P_{\text{AB}}(X)$). 

For path sampling schemes that do not evolve the shooting index from trial to trial, each new shooting index can be generated independently by directly drawing it from $p(k|X)$, implying that $p_{\text{sel}}(\hat k | X, k) = p(\hat k|X)$. In this case, the above expression reduces to
\begin{align}
    \rho(x|\text{SP}) = \int \text{d}X\, & P_{\text{AB}}(X) \sum_{k=1}^{L(X)} \biggl[ p(k|X) \notag \\
    &\times \sum_{\hat k=1}^{L(X)} p(\hat k|X) \delta(x - x_{\hat k}) \biggr]\, ,
\end{align}
where the sum over $k$ cancels and, renaming $\hat k$ as $k$, we arrive at:
\begin{align}
    \label{eq:sp_density_independent_drawing}
    \rho(x|\text{SP}) = \int \text{d}X\, P_{\text{AB}}(X) \sum_{k=1}^{L(X)} p(k | X) \delta(x - x_{k}) \, .
\end{align}

For a uniform shooting index distribution, $p(k|X) = 1/L(X)$, with independent drawing of shooting indices, as is done in standard TPS, we obtain the shooting point distribution
\begin{align}
    \label{eq:sp_density_TPS}
    \rho(x|\text{SP}) = \int \text{d}X\, P_{\text{AB}}(X) \frac{1}{L(X)} \sum_{k=1}^{L(X)} \delta(x - x_{ k})\, .
\end{align}
This distribution is very similar to the distribution $\rho(x | \text{TP})$ of points on transition pathways, but it differs from it by a weight factor proportional to $L(X)$. This factor arises because each transition path $X$ contributes $L(X)$ points to the distribution $\rho(x | \text{TP})$, but only one point to the shooting point distributions $\rho(x | \text{SP})$.

We can then relate the attempted shooting point distributions in Eqs.~\eqref{eq:sp_density_general},~\eqref{eq:sp_density_independent_drawing} and ~\eqref{eq:sp_density_TPS} to the extended space ensemble $P_\mathrm{AB}(X, k)$. For that, we determine the distribution of shooting points, not attempted, but as obtained from $P_\mathrm{AB}(X, k)$ by integrating/summing over pathways $X$ and shooting indices $k$:
\begin{align}
    \label{eq:k_configuration_density}
    \rho_{\text{ens}}(x) &= \int \text{d}X\, \sum_{k=1}^{L(X)} P_\mathrm{AB}(X, k) \delta(x - x_k) \notag\\
    &= \int \text{d}X\, \sum_{k=1}^{L(X)}P_{\text{AB}}(X) p(k|X) \delta(x - x_k) \notag\\
    &= \int \text{d}X\, P_{\text{AB}}(X) \sum_{k=1}^{L(X)} p(k|X) \delta(x - x_k)  \, .
\end{align}
This means that $\rho_{\text{ens}}(x)$, the distribution of points after summing over $k$, is the same as the distribution of attempted shooting points (Eq.~\eqref{eq:sp_density_independent_drawing}) one would obtain if shooting was performed by freshly drawing $k$ each trial from $p(k|X)$. Since $\rho_{\rm ens}(x) $ is independent of the specific shooting scheme yet depends on $p(k|X)$, we use it to assess the convergence of shooting schemes in the extended space $(X, k)$ as shown in Figs.~\ref{fig:fixed_length_TPS}~and~\ref{fig:flexible_length_TPS}.

\subsection{One-way and Two-way TPS \label{sec:TPS}}

We will next cast various TPS algorithms in the extended space formalism, starting with regular one-way and two-way shooting. In fixed- or flexible-length regular TPS, one draws the shooting index $k$ freshly at each round. As a result, there is no dependence of the picked shooting index on the previous shooting index $k^{\text{o}}$. In fact, since regular TPS is carried out in trajectory space only, the acceptance rules derived for the extended space do not directly apply to this case. Vice versa, casting the algorithm without modification to the extended space leads to irreversibility of the move (Fig.~\ref{fig:reversibility_issue}). However, we can construct an alternative TPS scheme that is reversible in the extended space, and closely resembles the original two-way shooting algorithm.

\begin{figure}
    \centering
    \includegraphics[width=0.7\linewidth]{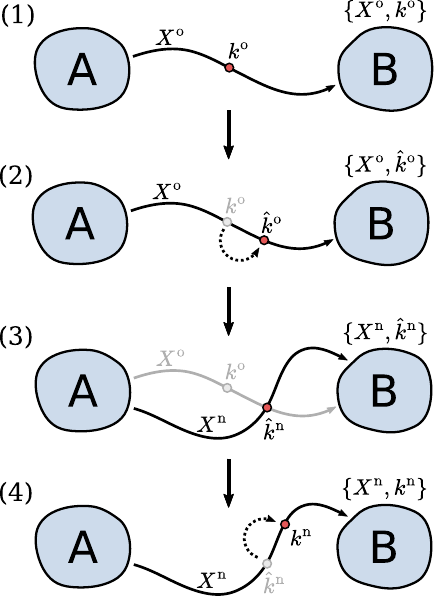}
    \caption{General scheme of a reversible shooting move in the extended space. The shooting index is shifted and the equations of motion are integrated. After that, the new shooting index is shifted a second time ensure reversibility.}
    \label{fig:shift_shoot_moves}
\end{figure}

We split the shooting move into three substeps as illustrated in Fig.~\ref{fig:shift_shoot_moves}. In the first part, the {\em index shift}, the shooting index is redrawn shifting the old shooting index $k^{\text{o}}$ to the point $\hat k^{\text{o}}$. In the second step, the shooting move itself is carried out from the point at $\hat k^{\text{o}}$ leading to the new path $X^{\text{n}}$. As the shooting move in most cases leads to a renumbering of points, we then define $\hat k^{\text{n}}$ as the index of the shooting point on the new path. As a last step, the shooting index is redrawn once again to ensure reversibility of the move. The whole shooting move results in a new path and shooting index $(X^{\text{n}}, k^{\text{n}})$.

Since we now work in the extended space, we need to specify the target shooting index probability $p(k|X)$ in order to obtain the full joint probability $P_\mathrm{AB}(X, k)$. To enforce a uniform shooting index distribution, we define $p(k | X)$ as:
\begin{align}
\label{eq:uniform_pk}
p(k | X) = \frac{1}{L(X)}\ \text{for}\ 1 \leq k \leq L \, .
\end{align}
For both index shift moves (steps (2) and (4) in Fig.~\ref{fig:shift_shoot_moves}), we draw the shooting index from the uniform distribution to simplify the acceptance and keep the algorithm close to standard TPS, such that
\begin{align}
p_{\text{sel}}(\hat k^{\text{o}} | X^{\text{o}})  = \frac{1}{L(X^{\text{o}})}\ \text{for}\ 1 \leq k \leq L(X^{\text{o}}) \notag\\ 
p_{\text{sel}}(k^{\text{n}} | X^{\text{n}} )  = \frac{1}{L(X^{\text{n}})}\ \text{for}\ 1 \leq k \leq L(X^{\text{n}}) \, .
\end{align}
Combining both selection probabilities leads to a symmetric index generation probability for the forward and the reverse move, $p_{\text{gen}}^\mathrm{fw} = p_{\text{gen}}^\mathrm{rv} = p_{\text{sel}}(\hat k^{\text{o}} | X^{\text{o}}) p_{\text{sel}}(k^{\text{n}} | X^{\text{n}})$. 
Therefore the acceptance probability based on Eqs.~\eqref{eq:acceptance_short_hasting} and~\eqref{eq:uniform_pk} is then given by 
\begin{align}
    \label{eq:acceptance_short_TPS}
    p_{\text{acc}}\bigl[ (X^{\text{o}}, k^{\text{o}}) &\to (X^{\text{n}}, k^{\text{n}}) \bigr] \notag\\
    &= H_{\text{AB}}(X^{\text{n}}) \min \biggl[ 1, 
    \frac{ p( k^{\text{n}} | X^{\text{n}}) }
    { p( k^{\text{o}} | X^{\text{o}})} \biggr] \notag\\
    &= H_{\text{AB}}(X^{\text{n}}) \min \biggl[ 1, 
    \frac{ L(X^{\text{o}}) }
    {  L(X^{\text{n}}) } \biggr] \, ,
\end{align}
which is identical to the acceptance criterion derived previously~\cite{Dellago2003c,Juraszek2006}. Note that for this algorithm the distribution of shooting points $\rho(x|\text{SP})$ follows Eq.~\eqref{eq:sp_density_TPS}.

In the analogous way, the shooting range algorithm~\cite{Jung2017} can be formulated in the extended ensemble. In this case, shooting indices $k$ are accepted according to a weighting function $w(r(x_k))$ along a collective variable $r(x)$, corresponding to a shooting index distribution:
\begin{align}
p(k | X) = \frac{w(r(x_k))}{\sum_i w(r(x_i))} \, .
\end{align}
In this case, the acceptance criterion becomes 
\begin{align}
    \label{eq:acceptance_shooring_range_TPS}
    p_{\text{acc}}\bigl[ (X^{\text{o}},  k^{\text{o}}) &\to (X^{\text{n}}, k^{\text{n}}) \bigr] \notag\\
    &= H_{\text{AB}}(X^{\text{n}}) \min \biggl[ 1, 
    \frac{ \sum_i w(r(x^\text{o}_i))  }
    { \sum_i w(r(x^\text{n}_i)) } \biggr] \, .
\end{align}
Setting $w(r)$ to unity if $a \leq r(x) \leq b$ with user defined bounds $a$ and $b$ and zero otherwise results in the original shooting range algorithm~\cite{Jung2017}.

\subsection{Fixed-length Aimless Shooting\label{sec:aimless_fixed}}

Aimless shooting with fixed path lengths is historically the basis for algorithms that evolve the shooting point from trial to trial. While there are different formulations of the algorithm in literature, we focus here on the two-point variant described in Ref.~\cite{Peters2010}. Here, we start with an initial path $X^{\text{o}}$ of fixed path length $L$. We assume an even path length in the following section, for odd path lengths integration steps need to be adjusted accordingly. The shooting index distribution is only nonzero at two points, $L/2 - \Delta k$ and $L/2 + \Delta k$, where $\Delta k$ is an adjustable parameter. This choice corresponds to the target shooting index distribution
\begin{align}
p(k | X) = \begin{cases}
\frac{1}{2} \ \text{if } k \in \{ \frac{L}{2} - \Delta k,\frac{L}{2} + \Delta k \},\\
0 \ \text{otherwise}
\end{cases} \, .
\label{eq:fixed_length_aimless_pk}
\end{align}
The two-point aimless shooting algorithm starts by selecting one of these two indices on the old path as the shooting index $\hat k^{\text{o}}$, which corresponds to the first index shift for regular TPS (Figure~\ref{fig:shift_shoot_moves}). For a positive shift, integration is performed in the backward direction by $L/2 + \Delta k - 1$ steps and in the forward direction by $L/2 - \Delta k$ steps to obtain the new path $X^{\text{n}}$ (vice versa for a backward shift). This operation produces a new path on which the index of the common point, i.e., the index $\hat k^{\text{n}}$ of the point where the old and the new path intersect, can be either at $L/2 - \Delta k$ or at $L/2 + \Delta k$. 

In principle, a second shift from $\hat k^{\text{n}}$ to $k^{\text{n}}$ would now be required to ensure that the move is reversible in the extended space. However, it is important to note that in \emph{fixed}-length aimless shooting the shooting index does not evolve, in contrast to to its flexible-length counterpart. While there is a chance that the next shooting move is initiated from the same \emph{point} as in the last trial, the shooting \emph{index} is always drawn independently as $L/2 - \Delta k$ or $L/2 + \Delta k$ each trial, effectively erasing the memory of the previous shooting index. In other word, given a certain path $X^{\text{o}}$, determining the new shooting index is always done relative to the center $L/2$ of the path and does not require knowledge of the previous shooting index. As a result, the procedure is Markovian in path space and therefore does not require an extended space formalism with second shift. We introduce a second shift as for standard TPS, which allows us to apply the extended space formalism and infer the distribution of the shooting points from which a shooting is attempted via Eq.~\ref{eq:sp_density_general}:
\begin{align}
    \label{eq:sp_density_aimless}
    \rho(x|\text{SP}) = \int \mathrm{d}X\, P_{\text{AB}}(X) \frac{1}{2}[& \delta(x_{L/2 - \Delta k} - x) \notag\\
    & +\delta(x_{L/2 + \Delta k} - x)  ] \, .
\end{align}
This is the distribution of points $\pm \Delta k$ away from the path centers. We validate this result via numerical simulations in section~\ref{result_section}. As for standard TPS in extended space, performing both shifts with symmetric probabilities leads to a cancellation in the acceptance probability
\begin{align}
    \label{eq:acceptance_short_mullen2015}
    p_{\text{acc}}\bigl[ (X^{\text{o}}, k^{\text{o}}) &\to (X^{\text{n}}, k^{\text{n}}) \bigr] \notag\\
    &= H_{\text{AB}}(X^{\text{n}}) \min \biggl[ 1, 
    \frac{ p(k^{\text{n}} | X^{\text{n}}) }
    { p(k^{\text{o}} | X^{\text{o}})} \biggr] \notag\\
    & = H_{\text{AB}}(X^{\text{n}}) \, .
\end{align}

\subsection{Flexible-length Aimless Shooting \label{sec:aimless}}

\begin{figure}
    \centering
    \includegraphics[width=0.7\linewidth]{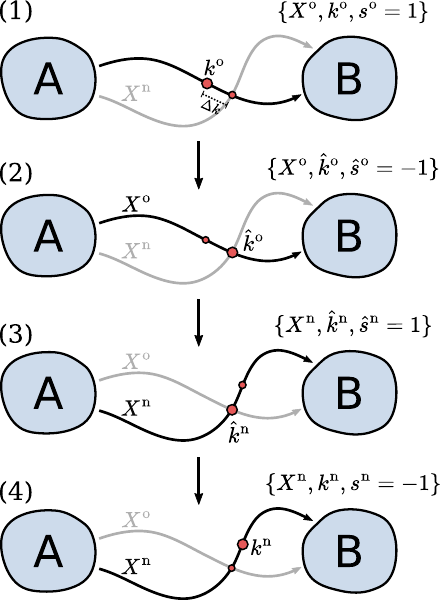}
    \caption{The two-point aimless shooting move in the extended space of paths $X$ and shooting indices $k$. (1) The move starts with the current path $X^{\text{o}}$, shooting index $k^{\text{o}}$ and direction $s^{\text{o}}$ that determines the position of the valid shooting indices. (2) The first index shift is performed by selecting one of the two potential shooting indices which generates $\hat k^{\text{o}}$. The direction $\hat s^{\text{o}}$ is flipped as the alternative shooting index is selected. (3) An extended space shooting move generates $X^{\text{n}}$ and $\hat k^{\text{n}}$. At the same time, a new direction $\hat s^{\text{n}}$ is drawn that determines if the second potential shooting point is on the forward or backward segment of the new path. (4) A second index shift is performed leading to $k^{\text{n}}$ with corresponding direction $s^{\text{n}}$. This shift is required to ensure the reverse move is possible (tracing bottom to top).}
    \label{fig:aimless_move}
\end{figure}

Flexible-length aimless shooting, as proposed by Mullen et al.~\cite{Mullen2015}, generalizes the aimless shooting move to a flexible-length scenario, which is particularly useful in systems with diffusive dynamics. Similar to fixed-length aimless shooting, the shooting index for the following trial is chosen from a narrow set of possible indices. Due to the flexible-length setting, these points are no longer defined to lie around the path center as in Eq.~\ref{eq:fixed_length_aimless_pk}, but they are close to the previous shooting point. Hence, in aimless shooting with flexible length pathways, there is true memory of the previous shooting point, making the treatment of the algorithm in the extended space formalism necessary. In the following, we will formulate the flexible-length aimless shooting in this formalism and will show that the acceptance probability include a ratio of path lengths, which was neglected in the original work~\cite{Mullen2015}. As demonstrated below numerically, this factor is necessary to sample the correct transition path ensemble.

As mentioned above, there is a crucial difference between the fixed-length and the flexible-length moves. Namely, in flexible-length algorithm, the new shooting index $k^{\text{n}}$ is selected depending on the previous index $k^{\text{o}}$ (see Fig.~\ref{fig:aimless_move}). In particular, the two-point variant described in~\cite{Mullen2015} samples the new shooting index from $\{k^{\text{o}}, k^{\text{o}} + s^\text{o}\Delta k\}$, where $s^\text{o} \in \{-1, 1\}$ defines whether the second shooting point is on the backward or the forward trajectory segment. Since the sign $s$ is preserved until a next trial is accepted, we include it in the extended ensemble $(X, k, s)$ (see SI for detailed balance derivation). To respect reversibility, the generation probability of $k^{\text{n}}$ is divided into two steps. We first choose a shooting index $\hat k^{\text{o}}$ on the old path according to
\begin{align}
\label{eq:aimless_pgen_A}
 p_{\text{gen}}( \hat k^{\text{o}} | X^{\text{o}}, k^{\text{o}}, s^{\text{o}}) = \begin{cases}
\frac{1}{2} \ \text{if } \hat k^{\text{o}} \in \{k^{\text{o}}, k^{\text{o}} + s^{\text{o}} \Delta k \},\\
0 \ \text{else}
\end{cases} \, .
\end{align}
If the opposing point, namely $k^{\text{o}} + s^{\text{o}} \Delta k$ is selected, we flip the direction $s^{\text{o}}$ such that the two possible shooting indices in the new state are preserved (see Fig~\ref{fig:aimless_move} at step 2). The new path $X^{\text{n}}$ is generated by shooting from $\hat k^{\text{o}}$ and integrating the equations of motion until a stable state is reached. We then obtain a new sign $\hat s^{\text{n}}$ by drawing from $\{-1, 1\}$ with equal probability. We assume $p(\hat s^{\text{n}}) = p_{\text{gen}}(\hat s^{\text{n}}) = \frac{1}{2}$.
Denoting $\hat k^{\text{n}}$ as the index of the shooting point on $X^{\text{n}}$, the new shooting index $k^{\text{n}}$ is generated according to:
\begin{align}
\label{eq:aimless_pgen_B}
 p_{\text{gen}}( k^{\text{n}} | &X^{\text{n}}, \hat k^{\text{n}}, \hat s^{\text{n}}) = \begin{cases}
\frac{1}{2} \ \text{if } k^{\text{n}} \in \{ \hat k^{\text{n}}, \hat k^{\text{n}} + \hat s^{\text{n}} \Delta k \},\\
0 \ \text{else}
\end{cases} \, .
\end{align}
Again, we obtain $s^{\text{n}}$ by flipping $\hat s^{\text{n}}$ if the opposing point was selected or else we leave it unmodified. In case $\hat k^{\text{o}}$ or $k^{\text{n}}$ fall outside of the old or new path, respectively, the move is rejected and the old path and shooting index are counted again.

This two-step procedure ensures that the path $X^{\text{o}}$ \emph{and} shooting index $k^{\text{o}}$ can be generated from $\{ X^{\text{n}}$, $k^{\text{n}} \}$ in the reverse move. 
We note that generation probabilities in Eqs.~\eqref{eq:aimless_pgen_A},~\eqref{eq:aimless_pgen_B} and $p_{\text{gen}}(s)$ are symmetric with probability $1/2$ and, as a result, they cancel in the acceptance criterion (Eq.~\ref{eq:acceptance_short}). In that case, given the analogous form of Eq.~\ref{eq:acceptance_short} derived in the SI, we obtain
\begin{align}
    \label{eq:acceptance_aimless}
    \frac{p_{\text{acc}}\bigl[ (X^{\text{o}}, k^{\text{o}}) \to (X^{\text{n}}, k^{\text{n}}) \bigr] }{p_{\text{acc}}\bigl[ (X^{\text{n}}, k^{\text{n}}) \to (X^{\text{o}}, k^{\text{o}}) \bigr]} =
     H_{\text{AB}}(X^{\text{n}})
    \frac{ p(k^{\text{n}} | X^{\text{n}}) }
    { p(k^{\text{o}} | X^{\text{o}}) }  \, .
\end{align}
Hence, we are left with the ratio of the shooting index distributions, which are not included in the original algorithm~\cite{Mullen2015}. As a result, accepting a path if $H_{\text{AB}}(X^{\text{n}}) = 1$ in a flexible-length setting violates detailed balance, which is also apparent in the results of numerical simulations (see section~\ref{result_section}). In the following, we choose a uniform shooting index distribution $p(k|X) = 1/L(X)$ as it requires no prior knowledge of the simulated system such as an order parameter. In that case, following the described two-step procedure, a properly weighted path ensemble can be obtained even \emph{a posteriori} by assigning each path a statistical weight of $\Omega(X) = 1 / L(X)$ in the case of a uniform shooting index distribution (see SI for derivation).

\subsection{Spring Shooting \label{sec:spring}}

\begin{figure}
    \centering
    \includegraphics[width=0.7\linewidth]{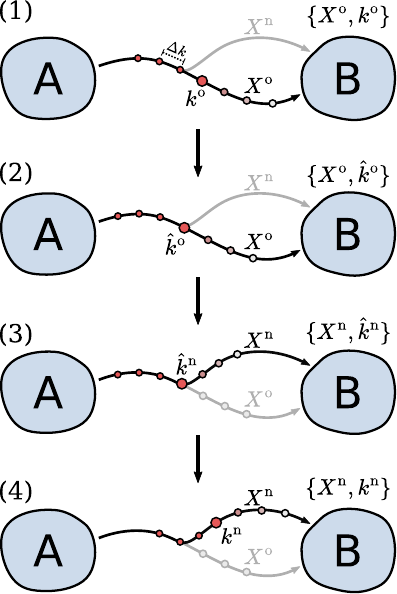}
    \caption{The spring shooting move in the extended space of paths $X$ and shooting indices $k$. (1) The move starts with the current path $X^{\text{o}}$ and shooting index $k^{\text{o}}$. (2) The first index shift generating $\hat k^{\text{o}}$ is performed by a weighted selection from neighboring points (illustrated using color gradient). (3) An extended space shooting move in the forward direction generates $X^{\text{n}}$ and $\hat k^{\text{n}}$. (4) A second index shift is performed leading to $k^{\text{n}}$. This shift is performed with opposite weights (see Eq.~\ref{eq:spring_shift_dist_inv}) and is required to ensure that the reverse move is possible (tracing bottom to top).}
    \label{fig:spring_move}
\end{figure}

Inspired by the flexible-length aimless shooting scheme, the spring shooting algorithm of Brotzakis et al.~\cite{Brotzakis2016} aimed to achieve a similar efficiency gain in a one-way shooting framework. In this scheme, the displacement $\Delta k$ from the previous shooting index is chosen according to an exponential distribution, i.e., a Boltzmann distribution based on a linear potential (constant force),  hence the scheme is named spring shooting. We next reformulate the algorithm in the shooting point MC framework (Fig.~\ref{fig:spring_move}), beginning by selecting the direction of the shooting move and assigning $s = -1$ for a forward shooting and $s=1$ for a backward shooting. Note that in this case $s$ is drawn freshly at the start of every trial, so it does not need to be included in the extended space. The displacement $\Delta k$ is then sampled according to:
\begin{align}
\pi(\Delta k) = \frac{\min[1, \exp(s\sigma \Delta k)]}{\sum_{\tau=-\Delta k_{\text{max}}}^{\Delta k_{\text{max}}}\min \bigl[ 1, \exp(s\sigma \tau) \bigr] }  \, ,
\end{align}
where $\sigma$ is the strength of the spring and $\Delta k_{\text{max}}$ is the maximum displacement. Following the notation of flexible-length aimless shooting, the shooting index $\hat k^{\text{o}}$ is set to $k^{\text{o}} + \Delta k$. The generation probability of the new shooting index becomes
\begin{align}
 p_{\text{gen}}( \hat k^{\text{o}} | X^{\text{o}}, k^{\text{o}}) = \begin{cases}
\pi(\hat k^{\text{o}} - k^{\text{o}}) \ \text{if } |\hat k^{\text{o}} - k^{\text{o}}| \leq \Delta k_{\text{max}},\\
0 \ \text{else}
\end{cases} \, .
\end{align}
Then, the new path $X^{\text{n}}$ is generated starting from the configuration at $\hat k^{\text{o}}$. The index of the common point of $X^{\text{n}}$ and $X^{\text{o}}$ on the new path is denoted as $\hat k^{\text{n}}$. As for aimless shooting, we then shift the new shooting index $\hat k^{\text{n}}$ a second time by sampling a shift $\Delta k$ from
\begin{align}
\label{eq:spring_shift_dist_inv}
\tilde \pi(\Delta k) = \frac{\min[1, \exp(-s\sigma \Delta k)]}{\sum_{\tau=-\Delta k_{\text{max}}}^{\Delta k_{\text{max}}}\min \bigl[ 1, \exp(-s\sigma \tau) \bigr] } \, ,
\end{align}
which corresponds to the following generation probability:
\begin{align}
 p_{\text{gen}}(  k^{\text{n}} | X^{\text{n}}, \hat k^{\text{n}}) = \begin{cases}
\tilde \pi(k^{\text{n}} - \hat k^{\text{n}}) \ \text{if } |k^{\text{n}} - \hat k^{\text{n}}| \leq \Delta k_{\text{max}},\\
0 \ \text{else}
\end{cases} \, .
\end{align}
This implies that, for a forward shot, first the shooting index is shifted preferably backwards. After generating the new path, the new shooting index is then generated by shifting, preferably forward. This ensures the move is reversible, and generation probabilities are symmetric (see SI). Consequently, the resulting acceptance ratio is identical to Eq.~\ref{eq:acceptance_aimless} and a distribution for the shooting index must be defined. 

In the original spring shooting algorithm proposed by Brotzakis et al.~\cite{Brotzakis2016}, the new shooting index $k^{\text{n}}$ is set implicitly to $\hat k^{\text{n}}$, the common point of the new and old path, i.e., the second shifting is omitted. While this formulation of spring shooting allows the reverse move in path space, i.e., regenerating the old path $X^{\text{o}}$ from the new path $X^{\text{n}}$, it does not regenerate the old shooting index $k^{\text{o}}$ from $k^{\text{n}}$. As a result, detailed balance is violated in the extended space framework. As for flexible-length aimless shooting, the ratio of shooting index distributions is not considered in the original form of the sampling scheme. For spring shooting, however, a reweighting factor for paths sampled without a defined shooting index distribution cannot be derived, due to the irreversibility of the shooting move in the extended space.

\begin{figure*}
    \centering
    \includegraphics[width=\textwidth]{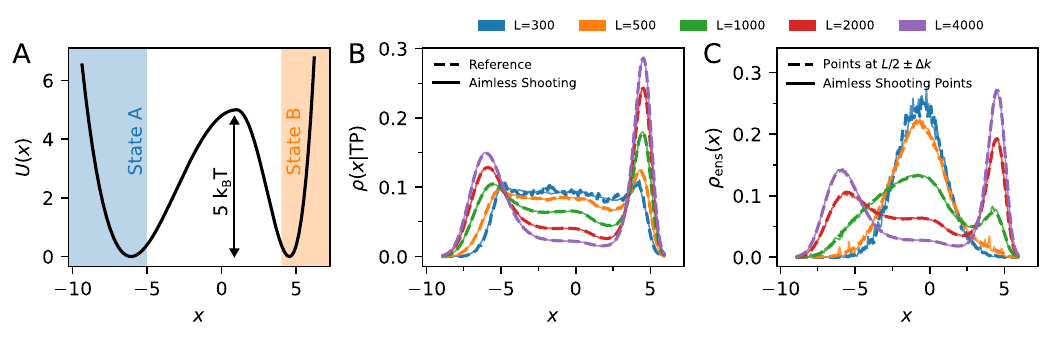}
    \caption{Fixed-length transition path sampling using aimless shooting. (A) The potential energy function of the model system and corresponding state definitions. (B) The density of points on transition paths for different fixed path lengths. Dashed lines correspond to the reference from a long equilibrium simulation, solid lines to the density obtained via aimless shooting. (C) Shooting point distributions for aimless shooting (solid line) and the points at $L/2 \pm \Delta k$ on transition paths from the equilibrium simulation (dashed lines).}
    \label{fig:fixed_length_TPS}
\end{figure*}

\begin{figure*}
    \centering
    \includegraphics[width=\textwidth]{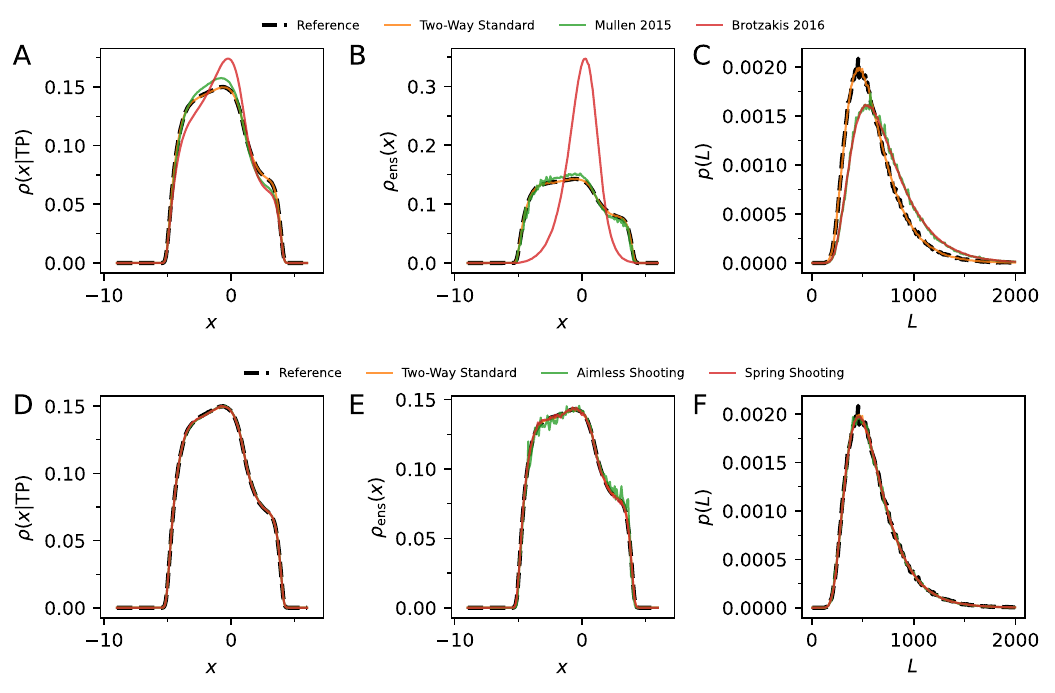}
    \caption{Flexible-length transition path sampling using aimless shooting as proposed by Mullen et al.~\cite{Mullen2015} and spring shooting as proposed by Brotzakis et al.~\cite{Brotzakis2016} (top row) and with the proposed corrections (bottom row). (A, D) Comparison of the density of points on paths between the equilibrium reference (dashed line) and path sampling algorithms. (B,E) Distribution of points after integrating out $k$ in the $(X, k)$ extended ensemble for the different path sampling algorithms. The reference distribution was obtained via reweighting the reference density in panel A with a factor proportional to $1/L(X)$. (C, F) The distribution of path lengths obtained using the different path sampling schemes.}
    \label{fig:flexible_length_TPS}
\end{figure*}

\section{Results and Discussion \label{result_section}}

As a reference for both fixed- and flexible-length path ensembles, we perform long equilibrium simulations from which we extract the transition path ensemble from spontaneous barrier crossings. All simulations are run employing overdamped Langevin dynamics in a one-dimensional, asymmetric double-well model (Fig.~\ref{fig:fixed_length_TPS}A) with a potential energy defined as (all simulation parameters in SI):
\begin{align}
    U(x) = \begin{cases}
0.2(x-1)^2 \bigl[0.01 (x-1)^2 - 1 \bigr] \ \text{if } x < 1,\\
0.2(x-1)^2 \bigl[0.16 (x-1)^2 - 4 \bigr] \ \text{else}
\end{cases} \, .
\end{align}
This barrier is about $5\,k_\text{B}T$ high. We chose an asymmetric barrier rather than a symmetric one to highlight differences in the obtained shooting point distributions.  

For the fixed-length path ensemble (Fig.~\ref{fig:fixed_length_TPS}B,C), we compare the density of points on paths $\rho(x | \text{TP})$ as in Eq.~\ref{eq:path_point_density} with the shooting point distribution as defined in Eq.~\ref{eq:sp_density_general}. The density of points in the transition region between states A and B decreases for longer paths as less time is spent in the barrier region. As a result, the points at $L/2 \pm \Delta k$ are less constrained to remain at the barrier top, as the time at which the path leaves state A becomes increasingly variable (Fig.~\ref{fig:fixed_length_TPS}C). For the longest paths in these simulations, there is hardly any difference between the density of points at $L/2 \pm \Delta k$ and $\rho(x | \text{TP})$. Since aimless shooting in a fixed-length setting samples these points as shooting points, the performance of the shooting scheme will depend on the chosen path length. For long paths, both uniform, two-way and aimless shooting initiate shooting moves effectively from the same shooting point distribution. As the shooting index is drawn independently at each trial and the shooting point distribution follows Eq.~\ref{eq:sp_density_aimless}, the efficiency of fixed-length aimless shooting arises from shooting close to path centers, rather than from any restoring force that directs shooting point toward the barrier top. For long path lengths, the acceptance probability of the shooting move decreases, as shooting points are more frequently drawn from the stable states rather than the barrier region. 

Within the flexible-length path ensemble, we observe that uniform two-way shooting converges to the reference ensemble, as indicated by overlapping $\rho(x | \text{TP})$ distributions and path length distributions $p(L)$ (Fig.~\ref{fig:flexible_length_TPS}). The density of points at $k$ denoted as $\rho_\text{ens}(x)$ matches the reweighted density of points on paths (Eq.~\ref{eq:sp_density_TPS}) obtained from the reference trajectory since shooting indices are selected with a uniform $p(k|X)$ on each path. In contrast, flexible-length aimless shooting as proposed by Mullen et al.~\cite{Mullen2015} and spring shooting as proposed by Brotzakis et al.~\cite{Brotzakis2016} do not converge to the reference, as shown in panels A and C of Fig.~\ref{fig:flexible_length_TPS}. Panel B shows that the shooting points for spring shooting are much more focused around the location of the barrier. While this is a desired feature, it clearly gives rise to the wrong distribution of transition paths (panels~A~and~C). We note that the deviations from the reference path ensemble are less apparent at higher barriers (see SI), where the distribution of shooting points is more localized for all methods. This is also the reason why spring shooting appeared to produce seemingly converged results in~\cite{Brotzakis2016}.

Including the shooting index distribution $p(k|X)$ in the acceptance probability and, for spring shooting, performing a second shift of the shooting index, recovers the convergence to the reference path ensemble (panels D, E, and F of Fig.~\ref{fig:flexible_length_TPS}). However, even though the methods differ in the way a new shooting index is drawn in each trial, shooting moves are now initiated from the same points as in standard uniform two-way shooting since we impose the same shooting index distribution. In addition, the small displacements of the shooting index in aimless and spring shooting introduce correlations between shooting points that are absent in uniform one- or two-way shooting. These correlations result in slow exploration of shooting points away from the barrier top (see SI). For flexible length aimless shooting and spring shooting, the shooting points were also generated disproportionately near the barrier regions. To maintain this desired focus, small displacements of the shooting points were performed in successive trials. This behavior is partially retained in the revised algorithms in the early stages of sampling, as the initial shooting index is often selected near the barrier top. However, with the imposed uniform shooting index distribution, the shooting index occasionally drifts away from the barrier to ensure full sampling the entire index distribution. This leads to low acceptance probabilities and extended sequences of rejected pathways, until the index returns to the barrier top by multiple small displacements in the right direction. The resulting persistent correlations considerably lower the efficiency of the algorithm. In standard TPS, on the other hand, the shooting index is redrawn independently at each trial, allowing it to move away from and return to the barrier instantaneously without such correlations.


\section{Conclusion}

We have proposed a theoretical framework to describe transition path sampling in an extended ensemble that includes paths and respective shooting indices. Revisiting algorithms that evolve the shooting point from trial to trial, we derived detailed balance equations for the extended ensemble, which led to corrections for aimless and spring shooting in flexible-length settings. While the deviation of path ensembles sampled via aimless shooting or spring shooting from the correct path ensemble is system-dependent, our experiments demonstrate a significant overlap in the distributions, suggesting that previous studies may be critically assessed and qualitative conclusions may remain largely valid. With recent advances in interface-based sampling schemes such as stone skipping~\cite{Riccardi2017} and wire fencing~\cite{Zhang2023} for transition interface sampling and virtual interface exchange~\cite{Brotzakis2019} for transition path sampling, we anticipate further development of shooting schemes for TPS that require minimal information about the system of interest.

\section*{Data Availability}

All data that support the findings of this study are available at \href{https://doi.org/10.5281/zenodo.14753554}{10.5281/zenodo.14753554}.

\begin{acknowledgments}
This research was funded in part by the Austrian Science Fund (FWF) 10.55776/F8100. For Open Access purposes, the author has applied a CC BY public copyright license to any author accepted manuscript version arising from this submission.
\end{acknowledgments}

\clearpage

\onecolumngrid
\normalsize
\patchcmd{\large}{15}{15}{}{}
\begin{center}
  \textbf{\LARGE Supplementary Information: Revisiting Shooting Point Monte Carlo Methods for Transition Path Sampling}\\[.2cm]
  Sebastian Falkner$^{1,2}$, Alessandro Coretti$^{1}$, Baron Peters$^{3,4}$, Peter G. Bolhuis$^{5}$, and Christoph Dellago$^{1,6,*}$\\[.1cm]
  {\itshape ${}^1$University of Vienna, Faculty of Physics, 1090 Vienna, Austria.\\
  \itshape ${}^2$Institute of Physics, University of Augsburg, Universit\"atsstraße 1, 86159 Augsburg, Germany.\\
  \itshape ${}^3$Chemical and Biomolecular Engineering, University of Illinois at Urbana-Champaign, Urbana, Illinois 61801, USA.\\
  \itshape ${}^4$Department of Chemistry, University of Illinois at Urbana-Champaign, Urbana, Illinois 61801, USA.\\
  \itshape ${}^5$Van ’t Hoff Institute for Molecular Sciences, University of Amsterdam, PO Box 94157, 1090GD Amsterdam, The Netherlands.\\
  \itshape ${}^6$ Research Platform on Accelerating Photoreaction Discovery (ViRAPID), University of Vienna, 1090 Vienna, Austria.}

  ${}^*$Electronic address: christoph.dellago@univie.ac.at\\
(Dated: \today)\\[2cm]
\end{center}

\setcounter{equation}{0}
\setcounter{figure}{0}
\setcounter{table}{0}
\setcounter{page}{1}
\setcounter{section}{0}
\renewcommand{\theequation}{S\arabic{equation}}
\renewcommand{\thefigure}{S\arabic{figure}}
\renewcommand{\thetable}{S\arabic{table}}
\renewcommand{\bibnumfmt}[1]{[S#1]}
\renewcommand{\citenumfont}[1]{S#1}
\renewcommand{\thesection}{S\Roman{section}}
\renewcommand{\thepage}{S\arabic{page}}

\titleformat*{\section}{\Large\bfseries}

\section{Simulation Details}

All simulations are performed in a one-dimensional model system with the potential energy $U(x)$ defined as:
\begin{align}
    U(x) = \begin{cases}
0.2(x-1)^2 \bigl[0.01 (x-1)^2 - 1 \bigr] \ \text{if } x < 1,\\
0.2(x-1)^2 \bigl[0.16 (x-1)^2 - 4 \bigr] \ \text{else}
\end{cases} \, .
\end{align}
We impose overdamped langevin dynamics with $k_\text{B}T = 1$ and propagate the system using
\begin{align}
    x_{i+1} = x_{i} + \Delta t D f(x) + \sqrt{2\Delta t D}\, \zeta \, ,
\end{align}
where $f(x)$ is the force at point $x$ and $\zeta$ is a random variable drawn from a standard normal distribution. For all simulations, we set $\Delta t D = 0.01$. For transition path sampling, we assume a point to be in state A for $x < -5$ and in state B for $x > 4$. We accumulate $24$ independent simulations each with $5 \times 10^5$ path sampling trials of which $10^3$ are discarded as equilibration period. Aimless and spring shooting are always performed with a maximum displacement of $\Delta k_\text{max} = 25$ and, for spring shooting, the spring strength is set to $\sigma = 0.1$.

\section{Extended Ensemble with additional Variables}

We can derive, analogous to the derivation of the Monte Carlo procedure for sampling $(X, k)$, a scheme for sampling the ensemble $(X, k, s)$, where $s$ may describe a sign for selecting a shooting index or even a direction to integrate the equations of motion (as for one-way shooting). First, we define an extended space as:
\begin{align}
Y = (X, k, s) \, .
\end{align}
In each step, a new path $X^{\text{n}}$, shooting point $k^{\text{n}}$ and direction $s$ is generated:
\begin{align}
(X^{\text{o}}, k^{\text{o}}, s^{\text{o}}) \to (X^{\text{n}}, k^{\text{n}}, s^{\text{n}}) \, .
\end{align}
with a generation probability 
\begin{align}
p_{\text{gen}}\bigl[ (X^{\text{o}}, k^{\text{o}}, s^{\text{o}}) \to (X^{\text{n}}, k^{\text{n}}, s^{\text{n}}) \bigr] \, .
\end{align}
As for the $(X, k)$ ensemble, we define $P(X, k, s)$ in such a way that marginalizing with respect to $k$ and $s$ should yield $P_{\text{AB}}(X)$:\begin{align}
\sum_k \sum_s P(X, k, s) = P_{\text{AB}}(X) \, .
\end{align}
This condition is fulfilled by:
\begin{align}
P(X, k, s) = P_{\text{AB}}(X) p(k | X) p(s) \, ,
\end{align}
where a sum over all $k$ and $s$ is guaranteed to lead to $P_{\text{AB}}(X)$ as marginal distribution. Detailed balance for a Markov chain Monte Carlo procedure that samples the joint probability $P(X, k, s) = P_{\text{AB}}(X) p(k | X) p(s)$ is given by:
\begin{align}
    P_{\text{AB}}(X^{\text{o}})  p(k^{\text{o}} | X^{\text{o}}) p(s^{\text{o}}) p_{\text{gen}}\bigl[ (X^{\text{o}}, k^{\text{o}}, s^{\text{o}}) \to (X^{\text{n}}, k^{\text{n}}, s^{\text{n}}) \bigr] p_{\text{acc}}\bigl[ (X^{\text{o}}, k^{\text{o}}, s^{\text{o}}) \to (X^{\text{n}}, k^{\text{n}}, s^{\text{n}}) \bigr] \\
    =  P_{\text{AB}}(X^{\text{n}}) p(k^{\text{n}} | X^{\text{n}})p(s^{\text{n}}) p_{\text{gen}}\bigl[ (X^{\text{n}}, k^{\text{n}}, s^{\text{n}}) \to (X^{\text{o}}, k^{\text{o}}, s^{\text{o}}) \bigr] p_{\text{acc}}\bigl[ (X^{\text{n}}, k^{\text{n}}, s^{\text{n}}) \to (X^{\text{o}}, k^{\text{o}}, s^{\text{o}}) \bigr] \, ,
\end{align}
yields
\begin{align}
    \label{eq:acceptance_full_SI}
    \frac{p_{\text{acc}}\bigl[ (X^{\text{o}}, k^{\text{o}}, s^{\text{o}}) \to (X^{\text{n}}, k^{\text{n}}, s^{\text{n}}) \bigr] }{p_{\text{acc}}\bigl[ (X^{\text{n}}, k^{\text{n}}, s^{\text{n}}) \to (X^{\text{o}}, k^{\text{o}}, s^{\text{o}}) \bigr]}
    = \frac{P_{\text{AB}}(X^{\text{n}})  p(k^{\text{n}} | X^{\text{n}}) p(s^{\text{n}}) p_{\text{gen}}\bigl[ (X^{\text{n}}, k^{\text{n}}, s^{\text{n}}) \to (X^{\text{o}}, k^{\text{o}}, s^{\text{o}}) \bigr] }{P_{\text{AB}}(X^{\text{o}})  p(k^{\text{o}} | X^{\text{o}}) p(s^{\text{o}}) p_{\text{gen}}\bigl[ (X^{\text{o}}, k^{\text{o}}, s^{\text{o}}) \to (X^{\text{n}}, k^{\text{n}}, s^{\text{n}}) \bigr]} \, .
\end{align}

\section{Reweighting Aimless Path Ensembles}

In the Main Text, we showed that the path ensemble resulting from flexible length aimless shooting as proposed in~\cite{SMullen2015} is not properly weighted. This is best illustrated in Figure 2a, where both the spring and aimless  algorithms do not recover the reference $p(x|TP)$ distributions.

Here, we show that this improper sampling can be easily corrected by introducing a simple reweighting factor for each of the generated paths in the ensemble. The proper detailed balance equation with symmetric generation probabilities is Eq.~(28) in the Main Text
\begin{align}
    \label{eq:acceptance_correct}
    \frac{p_{\text{acc}}\bigl[ (X^{\text{o}}, k^{\text{o}}) \to (X^{\text{n}}, k^{\text{n}}) \bigr] }{p_{\text{acc}}\bigl[ (X^{\text{n}}, k^{\text{n}}) \to (X^{\text{o}}, k^{\text{o}}) \bigr]} =
     H_{\text{AB}}(X^{\text{n}})
    \frac{ p(k^{\text{n}} | X^{\text{n}}) }
    { p(k^{\text{o}} | X^{\text{o}}) } \, .
\end{align}
However, the aimless shooting algorithm as proposed in~\cite{SMullen2015} uses a  different  acceptance 
\begin{align}
    \frac{p_{\text{acc}}\bigl[ (X^{\text{o}}, k^{\text{o}}) \to (X^{\text{n}}, k^{\text{n}}) \bigr] }{p_{\text{acc}}\bigl[ (X^{\text{n}}, k^{\text{n}}) \to (X^{\text{o}}, k^{\text{o}}) \bigr]} =
     H_{\text{AB}}(X^{\text{n}}) \, .
\end{align}
Clearly,  this would, in general, lead to different path statistics.
The simplest way to correct for this is to introduce a constant weighting factor $\Omega(X) $ for each sampled path. This would mean that we effectively  correct the aimless acceptance criterion as if we had used the following acceptance rule
\begin{align}
    \label{eq:acceptance_aimless_SI}
    \frac{p_{\text{acc}}\bigl[ (X^{\text{o}}, k^{\text{o}}) \to (X^{\text{n}}, k^{\text{n}}) \bigr] }{p_{\text{acc}}\bigl[ (X^{\text{n}}, k^{\text{n}}) \to (X^{\text{o}}, k^{\text{o}}) \bigr]} =
     H_{\text{AB}}(X^{\text{n}}) \frac{\Omega(X^{\rm n})}{\Omega(X^{\rm o})} \, .
\end{align}
The simplest way to ensure that this is identical to the proper acceptance rule in Eq.~\eqref{eq:acceptance_correct} is to set  the shooting point distribution $p(k|X)$ to
\begin{align}
    p(k|X) = \Omega(X) \, .
\end{align}
Since this should be true for all values of $k$, we require that the marginal of $P_\mathrm{AB}(X, k)$ in the reweighted flexible aimless path ensemble  will be equal to the marginal of $P_\mathrm{AB}(X, k)$  in the correct ensemble
That is,
\begin{align}
P_{\text{AB}}(X) = \sum_k P_{\text{AB}}(X) p(k | X) = \sum_k P_{\text{AB}}(X)  \Omega(X)  =  P_{\text{AB}}(X)  \Omega(X) L(X)  = P_{\text{AB}}(X)  \, .
\end{align}
This can only be achieved by  assigning each path the statistical weight
\begin{align}
\Omega(X) = \frac{1}{L(X)} \, .
\end{align} 
So, by reweighting each sampled path with a factor equal to the inverse of the length of the path, the bias in the path ensemble can be corrected. This is only possible when the two-step shifting of the shooting point is followed rigorously, and therefore, no reweighting factor can be derived for the original spring shooting as proposed in~\cite{SBrotzakis2016}. Still, when we use the corrected spring method described in the next section, and use the $1/L(X)$ reweighting factor, we do obtain the correct ensemble, as shown in the Main Text.\\

The above derivation starts with the premise that the flexible aimless algorithm samples the wrong distribution. However, the degree to which the distribution deviates depends on the maximum displacement $\Delta k_\text{max}$. We noticed that for the flexible-length aimless algorithm using a range of $\Delta k  \in [-\Delta k_\text{max},\Delta k_\text{max}]  \} $ the distribution of points on paths converges to the reference ensemble for large $\Delta k_\text{max}$, while it is still off for small  $\Delta k_\text{max}$.

To understand this, we construct the following argument. 
When doing a (small) shift in shooting point position as one would do in aimless shooting, one will (almost) always remain on the path itself (that is, the new shooting point lies on the path $X$), and one proceeds with the shooting move.
However, when $\Delta k_\text{max}$ becomes sufficiently large so that the new trial shooting point index falls outside of the path (of length $L(X)$), and is effectively in one of the stable states, the entire move has to be rejected. In that case one has to recount the current path in the Markov chain.

Assuming $\Delta k_\text{max}$  larger than the path length $L(X)$, we can argue that one is effectively choosing a random point on a stretch of the path of length $\Delta_p = 2 \Delta k_\text{max}+1$. If $\Delta k_\text{max}$ is sufficiently  large, the estimate for the chance of falling outside the (flexible) path for this random move in shooting point index is the $(\Delta_p-L(X))/\Delta_p = 1 - L(X)/\Delta_p$.
As one has to reject the move, the path is recounted and the  shooting point selection is repeated until one hits a trial shooting point within the length $L(X)$ region.  The probability to reject $n$ times in a row is $(1-L(X)/\Delta_p)^n$. 

The expected number of rejected moves can be computed as follows: 
Denoting $x\equiv 1-L(X)/\Delta_p$ the expected number of rejected moves is  
\begin{align}
      \langle n \rangle  =  \frac{\sum_{n=1}^\infty  n x^n}{  \sum_{n=1}^\infty  x^n}  \, .
\end{align}
The geometrical sums give
\begin{align}
 \langle n \rangle = \frac{\frac{1}{(1-x)^2}  }{ \frac{1}{1-x}} =  \frac{1}{1-x} = \frac{1} { L(X)/\Delta_p} =\frac{\Delta_p}{ L(X)} \, .
\end{align}
So in a proper sampling scheme, on average one has to recount the old path in the Markov chain $\Delta_p/L(X)$ times.  
As the $\Delta_p$ is just a constant, this gives a relative  weighting factor of $1/L(X)$, as expected. When one only has a small $\Delta_p$ or $\Delta k_\text{max}$, the rejection step almost never happens and all paths get the same weight.  
This explains why for small (maximum) displacement in the shooting point index the weighting of the paths are not taking into account correctly, while for larger displacement the paths ensemble converge to the correct distribution.

\section{Symmetric Generation Probabilities in Spring Shooting}

In the revised spring shooting algorithm, we first draw a displacement from the distribution:
\begin{align}
\pi(\Delta k) = \frac{\min[1, \exp(s\sigma \Delta k)]}{\sum_{\tau=-\Delta k_{\text{max}}}^{\Delta k_{\text{max}}}\min \bigl[ 1, \exp(s\sigma \tau) \bigr] } \, .
\end{align}
After generating the new path, we set $\hat k^{\text{n}}$ to the index of the shooting point on the new path. This index is then shifted a second time with a displacement given by:
\begin{align}
\tilde \pi(\Delta k) = \frac{\min[1, \exp(-s\sigma \Delta k)]}{\sum_{\tau=-\Delta k_{\text{max}}}^{\Delta k_{\text{max}}}\min \bigl[ 1, \exp(-s\sigma \tau) \bigr] } \, .
\end{align}
For given $\Delta k_1$ from $\pi(\Delta k)$ and $\Delta k_2$ from $\tilde \pi(\Delta k)$, the generation probability for the whole move is:
\begin{align}
 p_{\text{gen}}( k^{\text{n}} | X^{\text{n}}, X^{\text{o}}, k^{\text{o}}) = \pi(\Delta k_1) \tilde \pi(\Delta k_2) \, .
\end{align}
The generation probability of the reverse move is then given by:
\begin{align}
p_{\text{gen}}( k^{\text{o}} | X^{\text{o}}, X^{\text{n}}, k^{\text{n}}) = \pi(-\Delta k_2) \tilde \pi(-\Delta k_1)   \, ,
\end{align}
and the ratio of generation probabilities is then:
\begin{align}
 \frac{p_{\text{gen}}( k^{\text{n}} | X^{\text{n}}, X^{\text{o}}, k^{\text{o}})} 
 {p_{\text{gen}}( k^{\text{o}} | X^{\text{o}}, X^{\text{n}}, k^{\text{n}})} 
 =
 \frac{\pi(\Delta k_1) \tilde \pi(\Delta k_2)}{\pi(-\Delta k_2) \tilde \pi(-\Delta k_1)} \, .
\end{align}
Since $\pi$ and $\tilde \pi$ relate via
\begin{align}
 \pi(\Delta k) = \tilde \pi(-\Delta k) \, ,
\end{align}
the generation probabilities are symmetric
\begin{align}
 \frac{p_{\text{gen}}( k^{\text{n}} | X^{\text{n}}, X^{\text{o}}, k^{\text{o}})} 
 {p_{\text{gen}}( k^{\text{o}} | X^{\text{o}}, X^{\text{n}}, k^{\text{n}})} 
 = 1 \, .
\end{align}

\section{Convergence of Path Sampling Algorithms}

\begin{figure}
    \centering
    \includegraphics[width=\textwidth]{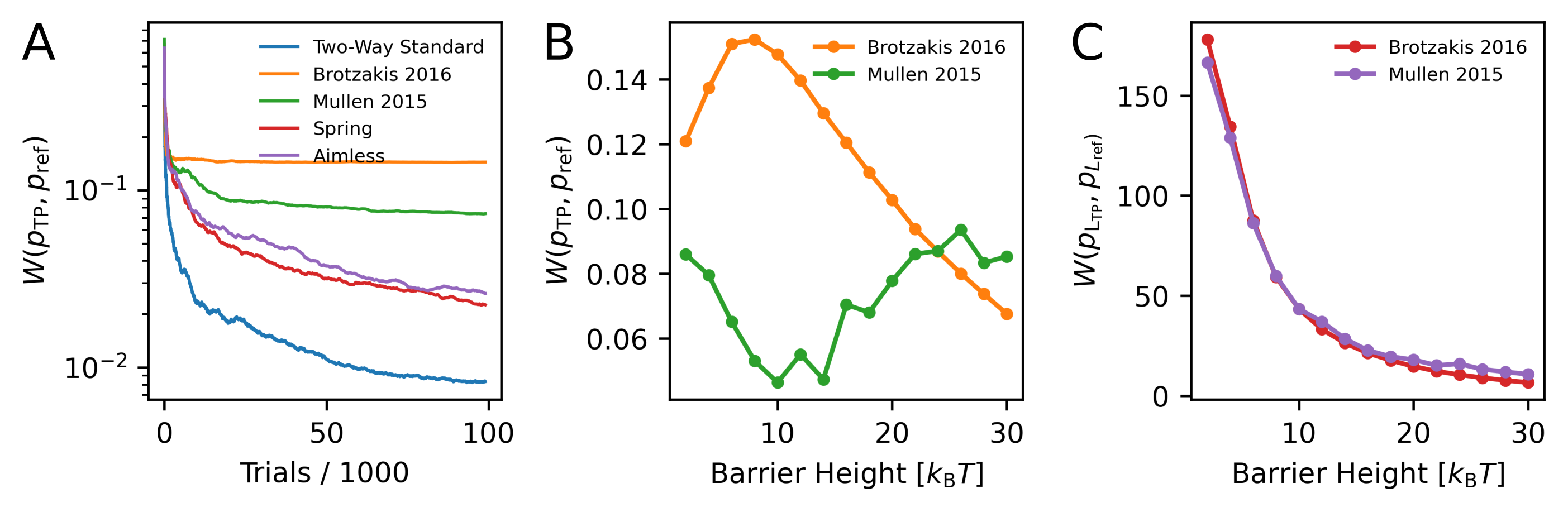}
    \caption{(A) Convergence of flexible-length transition path sampling algorithms including uniform two-way shooting, aimless shooting as proposed by Mullen et al.~\cite{SMullen2015}, spring shooting as proposed by Brotzakis et al.~\cite{SBrotzakis2016} and both with the proposed corrections. We use the Wasserstein distance between the reference distribution of points on paths $p_\mathrm{ref}(x|\mathrm{TP})$ and the distribution from each method at a specific trial $p_\mathrm{TP}(x|\mathrm{TP})$ averaged over $48$~runs. (B) Wasserstein distance between the distributions obtained from long simulations using the unrevised spring and aimless algorithms and the reference distribution as a function of the barrier height. Each point represents the average distance after $10^6$ trials from $192$~simulations. (C) Wasserstein distance with respect to the reference path length distribution for different barrier heights.}
    \label{fig:convergence_tests}
\end{figure}

In the main text, we show that the original spring and aimless algorithms do not converge to the correct distribution of points on paths. Here, we provide more details on the rate of convergence and the error of the unrevised methods with respect to the barrier height for the toy system (Fig.~\ref{fig:convergence_tests}). We measure convergence numerically using the Wasserstein distance between the reference distribution of points on the paths and the distribution generated by each method after a certain number of trials (Fig.~\ref{fig:convergence_tests}A). The convergence rate of the corrected methods is slower compared to regular two-way shooting with a uniform shooting index selection. This is, as stated in the main text, due to the introduction of correlations in the selection process of the new shooting point. Once the old shooting point $k^\mathrm{o}$ fluctuates away from the top of the barrier, the acceptance rate decreases drastically and the convergence is slowed. We see that the error of the original algorithms varies with the barrier height of the system (Fig.~\ref{fig:convergence_tests}B), where for spring shooting we see an initial increase in the error followed by a continuous decrease. For the original aimless algorithm, the error decreases initially with barrier height, however, at high barriers the error increases and the low acceptance rate leads to slow convergence. The error in the path length distribution continuously decreases with increasing barrier heights (Fig.~\ref{fig:convergence_tests}C), up to a point where the original algorithms produce an almost identical path length distribution as the corrected ones.

\section{Transition Path Sampling Algorithms}

Here, we provide pseudo-code for the TPS algorithms discussed in the main text.

\begin{algorithm}[H]
\begin{minipage}{0.5\textwidth}
        \centering
        \includegraphics[width=1\columnwidth]{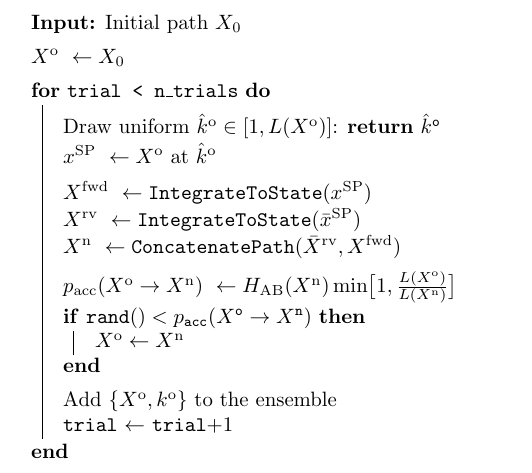}
    \end{minipage}\hfill
    \begin{minipage}{0.45\textwidth}
        \centering
        \includegraphics[width=1\columnwidth]{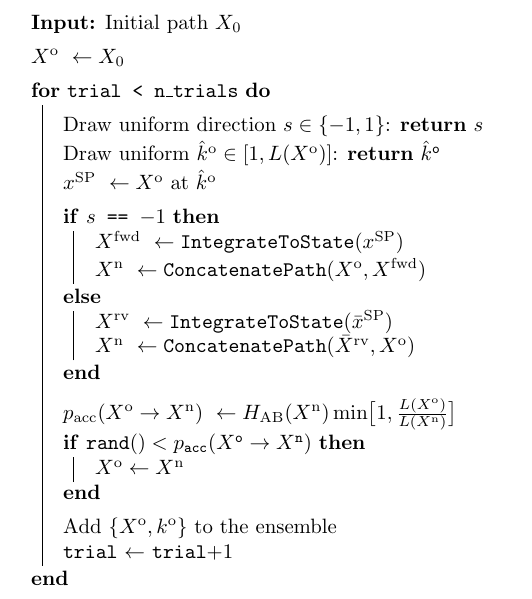}
    \end{minipage}
  \caption{Two-Way TPS (left) and one-Way TPS (right). The function \emph{IntegrateToState} generates a trajectory from the given point until a stable state is hit. The forward and reverse segments are fused by  \emph{ConcatenatePath}.}
  \label{alg:regular_TPS}
\end{algorithm}
\flushbottom \newpage 

\begin{algorithm}[H]
\begin{minipage}{0.42\textwidth}
        \centering
        \includegraphics[width=1\columnwidth]{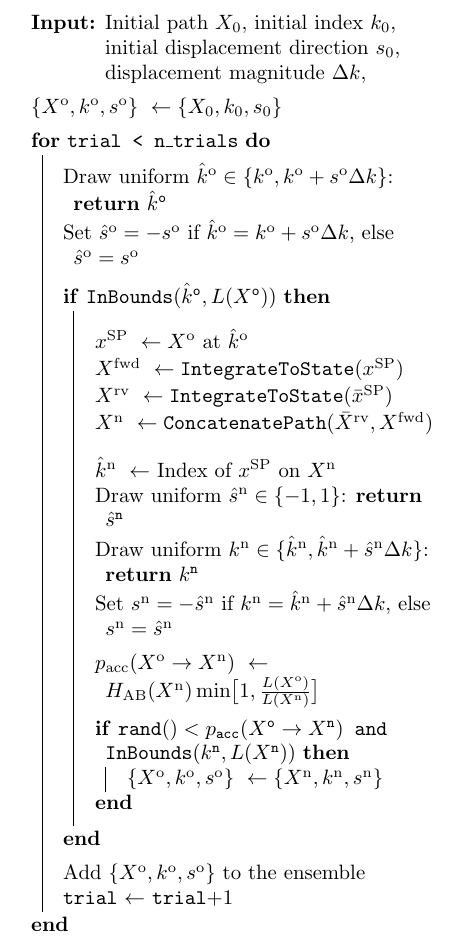}
    \end{minipage}\hfill
    \begin{minipage}{0.49\textwidth}
        \centering
        \includegraphics[width=1\columnwidth]{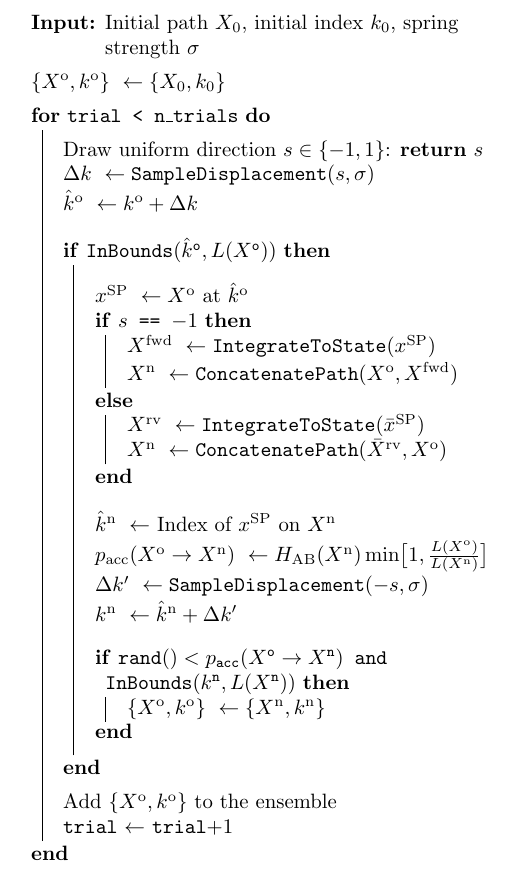}
    \end{minipage}
  \caption{Aimless shooting (left) and spring shooting (right) including the corrections from the extended space formalism. The function \emph{IntegrateToState} generates a trajectory from the given point until a stable state is hit. The forward and reverse segments are fused by  \emph{ConcatenatePath}. When a displacement is added to a shooting point index, we check if $1 \leq k \leq L(X)$ using \emph{InBounds}, otherwise the move is rejected. For spring shooting, \emph{SampleDisplacement} draws a $\Delta k$ according to a weight given by equation 30 in the main text.}
  \label{alg:aimless_spring_TPS}
\end{algorithm}

\end{document}